\documentclass[%
reprint,
prl,amsmath,amssymb,showkeys,floatfix
]{revtex4-2}

\usepackage{graphicx}
\usepackage{textcomp}
\usepackage{physics}

\usepackage{mathptmx}
\usepackage{microtype}

\renewcommand{\vec}[1]{\mathbf{#1}}
\newcommand{\EX}[1]{E_\text{#1}}
\newcommand{\ER}{\EX{R}}
\newcommand{\EL}{\EX{L}}
\newcommand{\Ex}{\EX{x}}
\newcommand{\Ey}{\EX{y}}

\begin{document}
\title{Generally applicable holographic  torque measurement \\
for optically trapped particles}
\author{Franziska Strasser$^\star$}
\author{Stephen M. Barnett$^\dagger$}
\author{Monika Ritsch-Marte$^\star$}
\author{Gregor Thalhammer$^\star$}\email{gregor.thalhammer@i-med.ac.at}
\affiliation{$^\star$ Medical University of Innsbruck, Institute of Biomedical Physics, Müllerstraße 44, 6020 Innsbruck, Austria}
\affiliation{$^\dagger$  University of Glasgow, School of Physics and Astronomy, Glasgow G12 8QQ, UK}
\date{\today}
\begin{abstract}
  We present a method to measure the optical torque applied to particles of arbitrary shape held in an optical trap, inferred from the change of angular momentum of light induced by the particle.  All torque components can be determined from a single interference pattern recorded by a camera in the back focal plane of a high-NA condenser collecting all forward scattered light. We derive explicit expressions mapping the measured complex field in this plane to the torque components. The required phase is retrieved by an iterative algorithm, using the known position of the optical traps as constraints.  The torque pertaining to individual particles is accessible, as well as separate spin or orbital parts of the total torque.
\end{abstract}

\keywords{optical angular momentum, optical torque, optical tweezers, phase retrieval, interferometry}

\maketitle

% Introduction and description of problem:
In recent years considerable interest has been directed towards understanding and utilizing the physics related to 'optical torque'. This refers to the torque exerted on microscopic particles in a light field, which is given by the rate of change in angular momentum transferred from the light field to the particles~\cite{friese1998optical}. Insightful experiments have been devised and carried out to quantify optical torque under various conditions~\cite{chang1985optical, nieminen_optical_2001, la_porta_optical_2004, parkin_measurement_2006, oroszi_direct_2006}. Yet, to date no generally applicable method is available to measure all components of the total torque---including spin as well as orbital contributions---acting on a trapped particle of arbitrary shape. In particular, the orbital contribution associated with the lateral optical phase profile of the light beam has proven challenging, as measuring the angular momentum of light requires the knowledge of the phase of the optical field \cite{crichton_measurable_2000,barnett2016natures}.  This becomes evident when considering the example of two Laguerre-Gauss beams of opposite chirality: Since they have the same intensity profile, the angular momentum of such beams cannot be deduced from the intensity, but rather from how the local wavefronts are twisted around the optical axis, which requires a phase measurement.

%describe our contribution: 
In this paper we present a calibration-free measurement approach which provides all components of the optical torque vector (containing spin and orbital parts) acting on arbitrary particles in optical tweezers~\cite{jones2015optical}. We derive the torque directly from light field: The change in angular momentum, and thus the torque, are determined with knowledge of phase and amplitude of the optical field in the back-focal plane (BFP)  of a high-NA condenser lens collecting all light in the forward direction. This approach has the great advantage that no calibration nor any \textit{a priori} information on the properties of the particle is required. To determine phase and amplitude we analyze the interference pattern of the light scattered by the particle and a reference field that emerges in the BFP. Specifically, for creating the reference field we use holographic optical tweezers~\cite{curtis2002dynamic, leach20043d} to generate one or more empty spots in addition to the spot occupied by a trapped micro-particle. From a single recorded interference pattern we recover the \textit{complex} light field in the BFP of the condenser lens by an iterative phase retrieval approach that was developed by us in the context of direct measurements of optical forces~\cite{thalhammer_direct_2015}.

Let us first explicitly derive the expressions for the torque based on the field in the BFP which are relevant for the above outlined procedure and which do not seem to be available in the literature. As the angular momentum density of light, which is given by the cross product of the position vector $\vec r$ with the momentum density of the electromagnetic field,
\begin{equation}
    \label{eq_angluar_momentum_density}
    \mathbf{j} = \varepsilon _0 \; \mathbf{r}\times (\mathbf{E}\times\mathbf{B})\, ,
\end{equation}
is conserved in free space, the torque $\vec T$ transferred to a particle can be determined from the integrated flux of angular momentum through a surface (with surface normals $\vec s$) encompassing the optically trapped object \cite{barnett_optical_2002, Jackson1975}. The  components of the torque are given by
\begin{equation}
    \label{eq:torque_general}
    T_i = - \int _S M_{il} \,\text{d}s_l \; ,
\end{equation}
where $M$ denotes the angular momentum flux density
\begin{equation}
    \label{eq:angular_momentum_flux_density}
    M_{il} =  \varepsilon_{ijk} r_j\left[\delta_{kl}\tfrac{1}{2}(\varepsilon_0 E^2+\mu _0^{-1} B^2)-\varepsilon_0 E_kE_l-\mu _0^{-1}B_kB_l\right]
\end{equation}
with $\delta_{kl}$ the Kronecker delta and $\epsilon_{ijk}$ the Levi-Civita symbol (using the Einstein summation convention).

In the far-field of a particle trapped by a monochromatic laser beam, the light it scatters travels essentially radially and can \textit{locally} be approximated by a plane wave with propagation direction $\vec k$. The radial components of $\vec E$ and $\vec B$, small but essential for angular momentum, can be expressed by gradients of the tangential components of $\vec E$ alone \cite{crichton_measurable_2000}, which considerably simplifies the explicit expressions.

\begin{figure}
  \includegraphics[width=0.95\columnwidth]{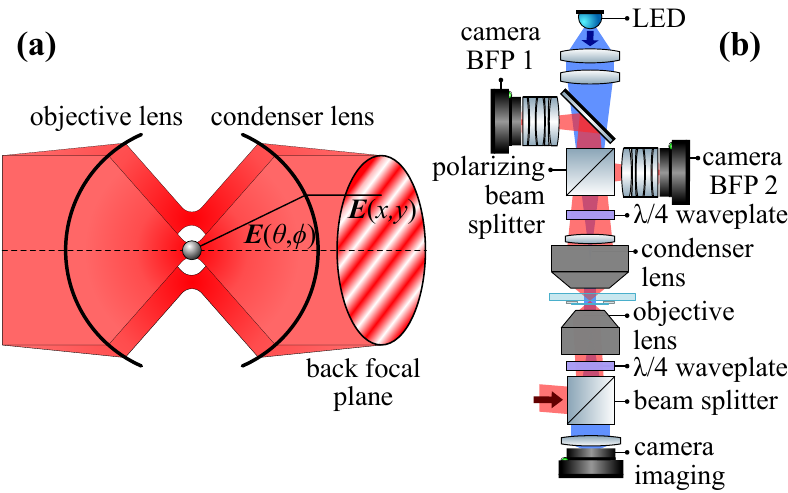}
  \caption{\label{fig:direct_torque_concept}
    (a) Schematic sketch of the electromagnetic field $E(x,y)$ arising from the interference of incident and scattered light, which is used to calculate the optical torque acting on the trapped particle. (b) Experimental setup.}
\end{figure}

We need to obtain expressions for the torque based on the field in the BFP, as this is what is accessible experimentally, i.e., we have to map the field between BFP and far-field, as sketched in Fig.~\ref{fig:direct_torque_concept}(a). For this we employ that the high-NA condenser lens obeys the Abbe sine condition: Light detected at a position $(x,y)$ in Cartesian coordinates is related to the scattering direction in spherical coordinate angles $(\theta, \phi)$ by $x = R \sin\theta \cos\phi$ and $y = R \sin\theta \sin\phi$, with $R$ denoting the maximum radius in the BFP where light can be detected, corresponding to light scattered at normal angle to the optical axis. For compactness we scale all spatial lengths in units of $R$ in the following, which makes the coordinates $x$ and $y$ dimensionless from now on. Furthermore, to conserve the energy flux in the mapping, the field amplitude in the far-field obtains an additional apodization factor of $\sqrt{\cos\theta}$ compared to the field in the BFP \cite{foreman2011computational}. % eq.(9) on p.343

We assume monochromatic electric fields described by $\vec E(x,y) e^{-i\omega t}$ and use the complex fields to calculate time-averaged quantities~\cite{barnett_optical_2002,crichton_measurable_2000}. In terms of the circular polarization components $\ER = \frac{\Ex + i \Ey}{\sqrt{2}}$ and $\EL = - \frac{\Ex - i \Ey}{\sqrt{2}}$ we obtain the following expressions for the components of the time-averaged torque $\vec T$,  as an integral
\begin{equation}
  T_i = -\frac{\varepsilon_0 c}{2\omega} R^2 \int_{A} \tau_i \,\text{d}x\,\text{d}y\label{eq:1}
\end{equation}
over the circular area $A$ of the pupil in the BFP with
\begin{align}
 \begin{split} 
   \tau_x &=
   \frac{x }{r^2}\left( 1 - \sqrt{1-r^2}\right) 
   \Delta_\text{LR} \\ 
   &\qquad
   -  \sqrt{1-r^2}\Im\left(\EL^\ast\partial_y\EL + \ER^\ast\partial_y\ER\right)
   %\text{d}x\,\text{d}y
 \end{split}
     \notag\\
 \begin{split}
   \tau_y &=
   \frac{y}{r^2} \left( 1 - \sqrt{1-r^2}\right)
   \Delta_\text{LR}\\ 
   &\qquad + \sqrt{1-r^2}\Im\left( \EL^\ast\partial_x\EL + \ER^\ast\partial_x\ER\right)
 \end{split}
     \notag\\
   \begin{split}
   \tau_z  &=
   %\iint
   \Delta_{\text{LR}} 
   + \Im \big[
   x (\EL^\ast\partial_y \EL + \ER^\ast\partial_y \ER)\\
   &\qquad\qquad\qquad -y  (\EL^\ast\partial_x \EL + \ER^\ast\partial_x \ER)
   \big]
   \\
   & =
       \Delta_{\text{LR}}
       +  \Im
       \big(
       r^2\EL^\ast \partial_\varphi \EL +
       r^2\ER^\ast \partial_\varphi \ER
       \big)
       \; . 
   \end{split} 
     \label{eq:torque_bfp}
\end{align}
Here  $\partial_x = \frac{\partial}{\partial x}$, and $(x, y)$ and $(r=\sqrt{x^2+y^2}, \varphi)$ are the normalized Cartesian and polar coordinates in the BFP, respectively, with $r\leq 1$. $\Delta_\text{LR}$ stands for $|\EL|^2-|\ER|^2$, which relates to the intensity difference $\Delta I_\text{LR}$ of the $L$ and $R$ component.

The use of the circular polarization components $\EL$ and $\ER$ instead of $E_x$ and $E_y$ is motivated by the fact  that Eq.~\eqref{eq:torque_bfp} is then independent of the relative phase between $E_L$ and $E_R$, i.e., no mixed terms in $E_L$ and $E_R$ occur.
The integration over the BFP of the condenser in Eq.~\eqref{eq:1} only gives the torque due to the total angular momentum flux in the forward direction, i.e., the outgoing flux. To obtain the exerted torque one needs to subtract the ingoing total flux $\vec T_{\text{in}}$, which can be obtained from measurements of the outgoing flux without particles.

%We have checked the validity of the expressions (\ref{eq:torque_bfp}) for simulated data against an alternative approach, where we express the electric field in the far-field by vector spherical harmonics, which are eigenstates of the total angular momentum operator $J_z$. This gives particularly simple expressions for the $z$-component of the angular momentum, see e.g. \cite[Eq.  (41)]{crichton_measurable_2000}. The other components can be obtained from this, e.g.~the $x$-component by first rotating the data on the far-field sphere around the $y$-axis by a right angle and then calculating the $z$-component of the rotated field.

Our experimental implementation for direct torque measurements is based on a holographic optical tweezers setup, which employs a liquid-crystal based spatial light modulator to control the trapping laser beams at 1064\,nm, and a water immersion objective lens (NA 1.2) for focussing, see Fig.~\ref{fig:direct_torque_concept}(b). For torque (and force) measurements we observe the light distribution in the BFP of a high-NA oil immersion condenser lens (Nikon, NA 1.4) with a digital camera (Sony IMX174 sensor) and a standard camera lens for both the right- and left-circular polarization components, which we separate using a quarter wave plate and a polarizing beamsplitter.
As in direct force measurements~\cite{smith_overstretching_1996, farre_force_2010, farre_optimized_2012, thalhammer_direct_2015} the condenser collects (nearly) all forward scattered light. This suffices, since $\mu$m-size cells or silica beads scatter only a small fraction (typically $<3\,\%$) in the backward direction due to their relatively small difference in refractive index compared to the surrounding buffer (water) \cite{farre_force_2010,thalhammer_direct_2015}.

To recover amplitude and phase of the field of the scattered light for each polarization component from the measured intensity image, which is a necessary input for Eq.~\eqref{eq:torque_bfp}, we employ common-path interferometry by adding a reference beam forming an empty trap. We deduce the complex light field (amplitude and phase) by solving for a field in the focal plane close to the trapped particle that, when numerically propagated to the measurement plane, matches the observed interference pattern of the trapping beams, using a gradient-based iterative optimization. Knowledge of the position of the traps imposed as a constraint waives the ambiguities associated with determining phase from a single image. For more details on the used algorithm and propagation model we refer to~\cite{strasser_direct_2021}.
\begin{figure}
  \centering
  \includegraphics[width=\columnwidth]{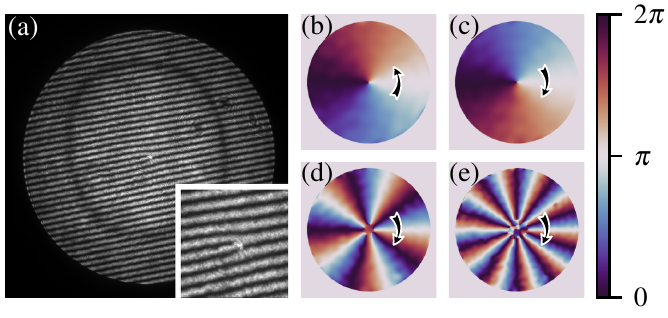}
  \caption{Results for field retrieval for beams with a spiral phase $\exp(il\varphi)$. (a) Observed interference pattern for $l=+1$, with inset showing the characteristic fork shape at the center. (b) Retrieved phase for $l=+1$, (c) for $l=-1$, (d) for $l=-5$, and (e) for $l=-10$.}
  \label{fig:spiral}
\end{figure}

%test for torque measurement
For a quality check of the outlined procedure for recovering the complex field for measurements of torque arising from orbital angular momentum, we use the SLM to imprint the well-known spiral phase pattern $\exp(il\varphi)$ to add orbital angular momentum to one of the trapping beams. The interference pattern in the BFP is recorded, in this example without any trapped particle. The results are shown in Fig.~\ref{fig:spiral}. The twist direction of the spiral phase pattern is indeed correctly recovered, and the root mean square difference between the reconstructed and the actual phase amounts to less than $2\,\%$ of $2\pi$. Similar reconstruction quality can also be achieved for larger azimutal index, with the exception of a small area around the central phase singularity, where the steep phase gradient diverts the light to regions in the focal plane that are outside the area where the field is reconstructed. 

\begin{figure}
  \includegraphics[width = \columnwidth]{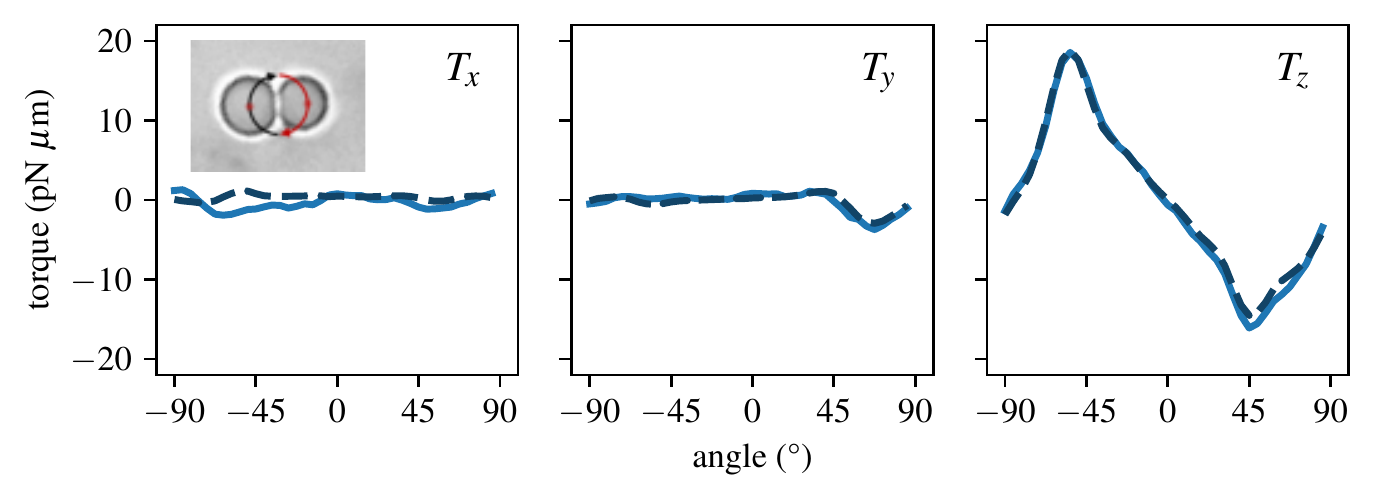}
  \caption{ \label{fig:J} Total optical torque applied to dumbbell system by two traps moved along a circular path shown in the inset. The torque from the simultaneous measurement (solid lines) is compared to the sequential 'force times lever' single trap measurement (dashed lines).}
\end{figure}

Having shown that we are able to accurately recover the phase, we proceed to optical torque measurements. As a test sample we use two silica microspheres ($d \sim$3\,\textmu{}m, see inset of Fig.~\ref{fig:J}), that touch each other and make up a dumbbell-shaped asymmetric object, which firmly adheres to a glass cover-slip. To exert an optical torque we use two focused spots at a distance of 3\,\textmu{}m, which we move in opposite directions along a circular path. We choose the center of the dumbbell object as the origin $\vec r_0$ relative to which the torque is defined. The region for the numerical phase retrieval in the focal plane covers both microspheres and beams, ie., both beams together are treated as a single beam with an asymmetric profile. 

This configuration was chosen because it allows us to determine the optical torque by an alternative way to check the validity of our approach: We repeat the experiment with only one focused spot switched on at a time and  sequentially measure the force $\vec F_i$ exerted by each beam. In this situation the torque for a single trap is well approximated by $(\vec r_i - \vec r_0) \times \vec F_i$ due to the force acting at a well localised position $\vec r_i$. From these data we calculate the total torque exerted in the original configuration with two beams as the sum of the individual measurements (with forces rescaled to account for different power of the trapping beams in the different configurations).

Fig.~\ref{fig:J} shows the result. Only the torque component $T_z$ along the axial direction is of considerable strength, as to be expected from the symmetry of the object and the placement of the beams. Some non-zero signal for $T_x$ and $T_y$ can be attributed to asymmetries induced by slight difference in the microsphere sizes and the placement of the beams. We see excellent agreement of the direct torque measurement with the sequential force based measurement. This confirms that our direct torque measurement method delivers correct results for a non-spherical particle and a complex beam shape.

% splitting J into L and S
Our experiments deliver the total torque applied to a trapped particle, given by the rate of exchange between the angular momentum of the particle and the field. Thus the question on how to (meaningfully) split the total angular momentum $\vec J = \vec S + \vec L$ into an internal spin part $\vec S$ and an external orbital part $\vec L$  is of relevance. This issue has been much discussed in the literature \cite{allen1992orbital,van1994spin,bliokh2014conservation,barnett2016natures} in past years. 
Splitting into a spin and an orbital part based on definitions involving unphysical non-transverse components in free space was found to be problematic, for instance concerning gauge invariance. Correcting this, by defining $\vec S$ and $\vec L$ in terms of strictly transverse fields (as for instance shown in \cite[Eq.~(17)]{barnett2016natures}), gives rise to two distinct quantities which are conserved \textit{separately}~\cite{van1994spin,van1994commutation,barnett2010rotation}. These two quantities do not represent true angular momenta by themselves---as opposed to their sum $\vec J$, which is a conserved quantity tied to the invariance of Maxwell's equations in free space under rotations. The so defined spin and orbital parts of the total angular momentum are separately accessible by suitable experiments and lead to individually measurable torque contributions in our system. 
\begin{figure}
  \includegraphics[width = \columnwidth]{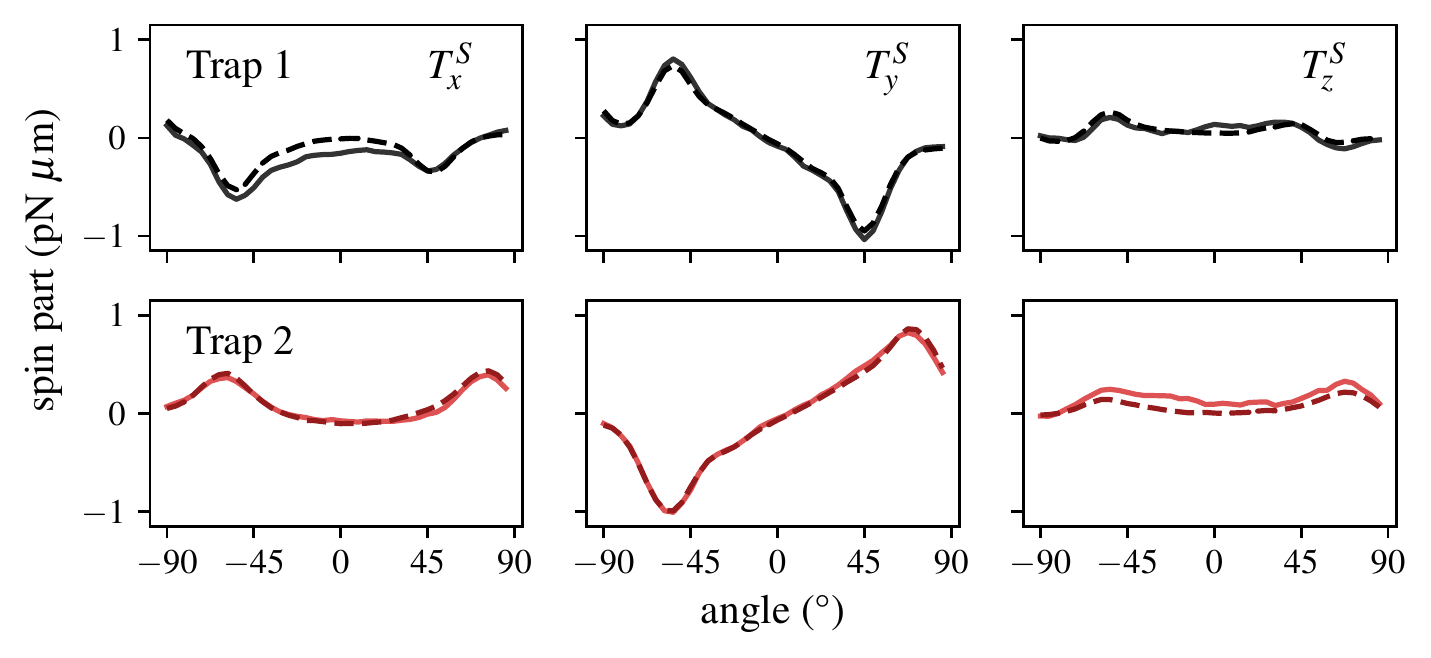}
  \caption{\label{fig:S} Simultaneously measured spin torque (solid lines) and sequentially measured torques with a single trap (dashed lines) for each microsphere of the dumbbell object.}
\end{figure}
Let us first look at the spin part. In the simple case of a plane wave with left circular polarization each photon carries a spin angular momentum of $\hbar$ with the spin vector pointing in the propagation direction, like the linear momentum. Since in the far-field of a localized source the light field propagates radially in a locally plane wave, we get the following simple expression for the torque $\vec{T}^\text{S}$ due to the spin part of the angular momentum flux
\begin{equation}
    \label{eq:spin_BFP_LR}
    \vec{T}^\text{S}   = - \frac{\epsilon_0 c }{2\omega} R^2  \int_A \left(x,y,\sqrt{1-r^2}\right)^T 
    \Delta_{\text{LR}}
    \dd{x}\dd{y}
    \; .
\end{equation}
It depends only on $\Delta_\text{LR}$, i.e.~on the \textit{intensity} difference of left and right circular polarization components in the BFP, and is therefore easier to measure than the orbital part. Changes of the polarization state contribute to the spin part of the torque. This effect is the underlying principle of methods using birefringent or asymmetric particles to apply optical torque in circularly polarized beams to induce rotations~\cite{parkin_measurement_2006}, also employed to quantify torques on DNA molecules~\cite{oroszi_direct_2006, pedaci2011excitable}. Particles such as cells or microspheres made of polystyrene or silica, however, hardly affect the polarization state, and thus---in our experimental situation---it is sufficient to use and detect a single polarization channel. Even without changing the polarization, an angular redistribution of circularly polarized light leads to a transfer of spin.

Fig.~\ref{fig:S} shows experimental results for the spin part defined in Eq.\eqref{eq:spin_BFP_LR} for each individual microsphere of the dumbbell object when both traps are turned on simultaneously (solid lines). The total spin torque on the system is small in the $x$ and $y$ direction. Hence we choose to show the individual spin torque for each trap emphasizing that we can distinguish the contributions by computationally propagating the individual localized light fields. Again, we find very good agreement with the sequentially measured spin torque where only a \emph{single} trap is turned on.

% orbital part
Main contributions to the orbital part $\vec T^L  = \vec T - \vec T^S$, cf.~Eqs.~\eqref{eq:torque_bfp} and \eqref{eq:spin_BFP_LR}, stem from terms containing derivatives of the field. Using the identity \( \Im \left(E^\ast\partial_x E\right) = |E|^2\partial_x\alpha\), one can also rewrite the expressions in terms of the gradient of the phase $\alpha$ of the complex field $E =|E| e^{i\alpha}$, in accordance with our previous statement that one needs to know the phase of the light field to determine its angular momentum.
In our setting of Fig.~\ref{fig:J}, %exemplary of the manipulation of extended asymmetric particles, 
the orbital part is much larger than the spin part due to the large effective lever length. The ability to measure modes with large orbital angular momentum with our method, as demonstrated by Fig.~\ref{fig:spiral}, is essential for this. Calculating ${\vec T}^L=\vec T - {\vec T}^S$ one can see that the orbital part also contains contributions from terms that are proportional to $\Delta_\text{LR}$ and thus depend on the polarization state (unless for linear polarization where $\Delta_\text{LR}=0$).
This is a consequence of the enforced transversality when splitting the angular momentum into a spin and an orbital part. A good example for this is a focussed beam where the inclination angle of the local $\vec k$-vectors with respect to the optical axis gives rise to a 'spin-orbit conversion'~\cite{bliokh2010angular,ruffner_optical_2012,zhao2007spin}.  One can directly see that strong spin-orbit conversion is specific to high-NA optics, e.g., from a term  $\left(1-\sqrt{1-r^2}\right)\Delta I_\text{LR}$ occurring in the $z$-component of ${\vec T}^L$, which only contributes for $r\approx 1$, i.e., for large angles $\theta$ from the optical axis. 

To summarize, we have presented a method to directly measure all components of the torque exerted on an optically trapped particle. We obtain high-resolution information about phase, amplitude and polarization of the transmitted light in the far-field by imaging the interference pattern in the back focal plane with a digital camera for two polarization modes and by numerically reconstructing the field from a single image. From this complete information about the field in the BFP we calculate \textit{all components} of the torque, which we can further split in contributions from spin and orbital parts of the angular momentum, providing deeper insights in the mechanisms for inducing torque. 

Our method fully incorporates the more intricate conditions of high-NA focusing and imaging, which is typical for optical trapping, and which leads to puzzling effects such as polarization dependent contributions to the orbital part of angular momentum.
Our calibration-free approach is applicable in general settings and incorporates special cases, where, e.g., only the change in polarization is detected (for torque induced mostly by spin angular momentum \cite{nieminen_optical_2001}), or only a few spatial modes with different orbital angular momentum are analyzed (for direct measurement of changes of orbital angular momentum \cite{Parkin2004}). Recovering the phase from a single interference pattern enables applications with dynamic settings, but needs high computing power. Data analysis in real-time appears feasible with optimized implementations. 

It is well suited to measure torques for asymmetric particles with unknown properties, which is a common setting in the life sciences. The complete, high-resolution information about the light field allows us to extend torque measurements to an ensemble of simultaneously trapped particles, delivering torque measurements for each particle individually. We hope that it will find many uses, both for studying fundamental aspects of optically induced mechanical effects and for applications, especially  in biology, measuring torque on the cellular level.

\begin{acknowledgments}
This work was supported by the Austrian Science Fund (FWF) (P29936-N36) and by the Royal Society (RP 150122).
\end{acknowledgments}

\appendix

% \bibliography{references}


\end{document}